\documentclass[a4paper,12pt]{article}
\usepackage[utf8]{inputenc}
\usepackage{epsfig}
\usepackage{amsmath,amssymb}
\usepackage{textcomp}

\newcommand{\be}{\begin{equation}}
\newcommand{\ee}{\end{equation}}
\newcommand{\bea}{\begin{eqnarray}}
\newcommand{\eea}{\end{eqnarray}}
\newcommand{\bv}{\bar{v}}

\title{Physical process first law and caustic avoidance for Rindler horizon }
\author{Srijit Bhattacharjee\footnote{email: srijitb@iitgn.ac.in}, Sudipta Sarkar\footnote{sudiptas@iitgn.ac.in} \\
Indian Institute of Technology Gandhinagar\\ Ahmedabad, Gujarat- 382424, India.}
\date{}
\begin{document}
\maketitle
\begin{abstract}
We study the perturbation induced by a slowly rotating massive object as it passes through a Rindler horizon. It is shown that the passage of this object
can be approximately modeled as Delta\,function type tidal distortions hitting the horizon. Further, following the
analysis presented by Amsel, Marolf and Virmani related to the issue of the validity of physical process first law, we establish a condition on the size of
the object so that this law holds for the Rindler horizon. 
\end{abstract}
\section{Introduction}
Black holes are arguably the simplest systems (macroscopically) in our Universe as they are characterized by only few parameters namely mass ($M$), charge
($Q$), and angular momentum ($J$). Yet many of their features are not understood to the level of complete satisfaction. Most of the interesting properties
of black holes have emerged due to the presence of the ``event horizon'' which acts as a causal boundary. Striking resemblance between thermodynamic
parameters of a system and the area, surface gravity etc. of a black hole event horizon has been an active area of investigation since the laws of black
hole mechanics were first put forward \cite{Bekenstein, BCH}. It was also emphasized that laws of black hole thermodynamics can be extended to more general
settings which are collectively asserted as ``causal horizons'' \cite{JP}. Causal horizons are the boundary of the past of any timelike curve of infinite
proper length in the future direction. Particularly they include Rindler horizon perceived by an uniformly accelerated observer. Although the laws of black hole thermodynamics apply quite well to all the causal horizons but the Rindler (RH) appeared to be an exception as argued in \cite{JP}. This disagreement happens in the context of ``physical process version of the first law'' (henceforth will be called as PPFL).\\
 This version of the first law describes dynamical change in horizon area in response to a flux of stress energy through the generators of the horizon
\cite{HH, CarterReview, Wald}. The PPFL for black holes can be written in the form

\be
 \label{ppfl}\frac{\kappa}{8\pi}\Delta A_H=\int_{\mathcal{H}} T_{\mu\nu}\chi^{\mu} d\Sigma^{\nu}
 \ee

where, $T_{\mu\nu}$ is the stress energy flux that is crossing the horizon and $\chi^\mu$ is the horizon generating Killing vector. Now, to derive the PPFL one
has to integrate the Raychaudhuri equation for null geodesic congruences \cite{Wald}. A crucial assumption in this derivation is that the process must be
sufficiently stationary in the sense the expansion and shear are weak enough to neglect the higher order terms along each generators. This approximation is justified as long as there is no formation of caustic.\\
Now for RH this seemed to be invalid. Consider passage of a planet of mass $m$ across the horizon of a Schwarzschild black hole of mass $M$. The planet started it's journey from an infinite distance from the Schwarzschild horizon and falling freely. This situation has been analyzed in \cite{PRS, PTM} within the membrane framework. It was shown that the process will always be quasi-stationary or no caustic will form only if $r \gg m\,M$ , where $r$ is the radius of the planet. Considering Rindler horizon as the $M\rightarrow \infty$ limit of Schwarzschild black hole would therefore seem to indicate that the planet must be of infinite radius however small it's mass may be! Thus for any finite sized object PPFL will not hold as the process can not be made quasi-stationary. But, in \cite{AMV}, it was clarified that the above argument is not completely acceptable because in the analysis of Price et al. \cite{PTM} the planet under consideration had started it's journey from rest at infinite distance apart from the 
black hole horizon which doesn't have smooth limit for the case of Rindler horizon. With the help of a more refined treatment and considering a local characterization
of formation of caustics it was shown in \cite{AMV} that the PPFL for RH holds non-trivially, in fact it holds for any general bifurcate Killing horizon.

To establish their claim, Ref. \cite{AMV} considered passage of a weakly self-gravitating object of mass $m$ passing through a RH in flat spacetime. Since
the derivation of PPFL traces generators of the horizon only back to the unperturbed past horizon, one only has to make sure no caustic forms to the future
of the bifurcation surface. In \cite{AMV}, it was shown one can indeed avoid the formation of caustics in the region of interest and the first law holds for
Rindler horizon. Moreover, there is a bound on the size of the object given by the relation $r \gtrsim \sqrt{E_\chi/\kappa}$ for which caustic will not
form and the process will remain quasi stationary, where $E_\chi=m\kappa z_0$ and $\kappa$ are the Killing energy of the incident object and the surface
gravity defined by the Killing field $\chi$ respectively. The object falls from a height $z_0$ from the horizon. \\

In this article we have considered two such physical processes. First we have analyzed a charged non-rotating object passing through the RH and determine the bound on the size of the object to have the PPFL holds in the region of interest. Next we have considered the case when a slowly {\textit rotating} object of mass $m$ falls across the Rindler horizon from rest. The primary
goal of these investigations is to generalize the analysis depicted in \cite{AMV}. For the rotating case, the horizon is now imparted by angular momentum of the
falling object on top of it's mass. This is analyzed by reducing it to an equivalent problem where the horizon is being hit by three tidal pulses at three
different moments. It will be shown in section \ref{sec:Kerr}, PPFL holds and a bound on the size of the object can be
established for which the quasi stationarity is maintained throughout the future of the bifurcation surface.

\section{PPFL and condition for caustic}
\label{sppfl}

To measure dynamical changes of any surface which is generated by congruences associated with any vector field we need to know how the expansion, shear
etc. changes along the congruences. Hence we must refer to the Raychaudhuri equation to get these information. In fact, the starting point of derivation of
the PPFL is the Raychaudhuri equation for null congruences on any bifurcate Killing horizon. Here we briefly outline the essential steps of the derivation.
It is convenient to parameterize the geodesics by Killing time $v$ instead of the affine parameter $\lambda$. Consequently $v \rightarrow -\infty$ now
corresponds to the bifurcation surface and $v\rightarrow \infty$ corresponds to the asymptotic future. We consider a change in the stationary configuration
of the  horizon induced by some flux of energy through it which will perturb it for a brief interval of time and will settle down again to an equilibrium
state. Then the equation governing the change in expansion $\theta$ for null congruences associated with horizon-generating Killing vector $\chi$ becomes,

\be
 \frac{d \theta }{d v} = \kappa \theta - \frac{ \theta^2}{2} - \sigma_{\mu \nu} \sigma^{\mu \nu} -8\pi T_{\lambda \sigma} \chi^{\lambda} \chi^{\sigma},
\label{thetaeqn}
\ee

where we have used Einstein equation and $\kappa$ is the surface gravity of the background Killing horizon defined by $\chi^\nu \nabla_\nu \chi^\mu = \kappa \chi^\mu$. Now we specialize to
a situation where the change induced by the process to the stationary background is small. This means in (\ref{thetaeqn}), we can take the expansion and
shear to be weak enough so that we may ignore the terms beyond linear order in $\theta$ and $\sigma$ and obtain,

\be
 - \frac{d \theta}{dv} + \kappa \theta = 8\pi
T_{\lambda \sigma} \chi^{\lambda} \chi^{\sigma}=S(v), \label{flaw}
 \ee

  From (\ref{flaw}), we solve for expansion $\theta$ by employing Green's function technique and future stationary boundary condition to obtain,

 \be
 \theta(v) = \int e^{\kappa (v - v')} S(v') dv'. \label{thetaint}
 \ee

 Now recalling the fact that the expansion $\theta$ measures the fractional rate of increase of the cross-sectional area element of a bundle of null generators
 over a finite range of Killing time, we may rewrite
(\ref{flaw}) in the desired form:

\be
\frac{d(\Delta A)}{dA} = \int_{-\infty}^{\infty} \theta \: dv = \int_{-\infty}^{\infty} dv \int_{v}^{\infty} dv' e^{\kappa (v-v')} S(v'). \label{ppfl}
\ee

 Integrating the expression (\ref{ppfl}) over $v$ we get the PPFL:

 \be
 \frac{d(\Delta A)}{dA} = \frac{8 \pi }{\kappa}
\int_{-\infty}^{\infty} dv \ T_{\mu \nu} \chi^{\mu} \chi^{\nu} \, .\label{ppfl1}
 \ee

 Since the r.h.s. of (\ref{ppfl1}) signifies amount of matter energy
crossing the horizon we can recast it as familiar form of first law:

\be \label{fflaw} \frac{\kappa \Delta A}{8 \pi } = \Delta E_\chi.
 \ee

 Note that, the derivation of PPFL from (\ref{thetaeqn}) crucially depends on the
fact that the process remains quasi-stationary which means to the future of the bifurcation surface of the background stationary space time both shear $\sigma_{\mu\nu}$ and expansion $\theta$ remain weak enough such that we can ignore all second order terms in the Raychaudhuri equation. We can actually estimate the threshold value of
$\theta$ when the process will fail to remain quasi-stationary. It is easily observed from (\ref{thetaeqn}) if $\frac{\theta }{2 \kappa} \sim 1$ then the
second order term becomes relevant. This can be observed from the null Raychaudhuri equation (\ref{thetaeqn}), solving it for $\sigma^2= 0 = S(v)$ gives;

\bea
\bar \theta(v) = \frac{1}{1+\left( \bar
\theta_0^{-1}-1\right)e^{\kappa(v_0 - v)}} \,,
\eea

where $\bar \theta = \frac{\theta}{2\kappa}$ and $\bar
\theta(v_0) = \bar \theta_0$. Clearly when $\bar \theta_0 > 1$, then $\bar \theta$ increases as we move along past and diverges at some finite Killing time before the bifurcation surface. Therefore, if for any cross section $v = v_0$, the horizon expansion is strong enough so as to satisfy

\be \theta \ge 2\kappa \label{condn} \ee

then the Raychaudhuri equation implies that a caustic will be developed at some finite $v < v_0$. In the following section, this condition will
be used to get a bound on the size of the object which crosses the horizon.\\
 To analyze the distortion induced by matter flux at the horizon one needs to
include the evolution of horizon shear also. This will be apparent after we write down the ``tidal force equation'',

\be \frac{d \sigma_{\mu \nu}}{d v} = \left( \kappa - \theta\right) \sigma_{\mu \nu}- \sigma_{\mu \sigma} \sigma^{\sigma}{}_{\nu}+ \frac{ \sigma^2}{2}Q_{\mu
\nu} + \left( 2 \sigma_{\mu \sigma} + \theta Q_{\mu \sigma}\right) \sigma^{\sigma}{}_{\nu}- \mathcal{E}_{\mu \nu}. \label{tidaleq}
\ee

where $\mathcal{E}_{\mu \nu} := Q^{\alpha}{}_{\mu} Q^{\beta}{}_{\nu} C_{\alpha \lambda \beta \sigma} \chi^{\lambda} \chi^{\sigma} $ is the electric part of
the Weyl tensor $C_{\alpha \lambda \beta \sigma}$ and $Q^\mu_\nu $ is the projection operator onto the space like cross section of the horizon. The
projector $Q_{\mu\nu}$ is defined as:

\be Q_{\mu \nu} = g_{\mu \nu} + \chi_\mu l_\nu+ \chi_\nu l_\mu
 \ee

 where $l^\mu$ is the second null vector satisfying $l^\mu \chi_\mu = -1$. This second null
vector is needed to uniquely determine the orthogonal subspace spanned by the vectors which are normal to both $l^\mu$ and $\chi^\mu$. In the following we
will consider two situations. First, when the generators of the horizon will be crossed by the stress energy tensor of the external matter and second is
when it will not be crossed. For the second case, one doesn't have the source term in the right hand side of
(\ref{thetaeqn}), but one could have nonzero shear due to presence of the $\mathcal{E}_{\mu \nu}$ term in the tidal force equation and this shear in turn
acts as a source for the $\theta$ equation. This is why we need the tidal force equation. If we consider the perturbation to be of order $\epsilon$, a
small dimensionless parameter then $\mathcal{E}_{\mu \nu}$, $\theta$ and $\sigma$ are also of the same order. Now when the perturbation is absent across
the horizon we still get a nonzero shear by solving the tidal force equation retaining terms up to first order in $\epsilon$. From (\ref{thetaeqn}) it is
clear that in this case $\theta$ equation has to be of order $\epsilon^2$ due to the presence of the term $\sigma^2$. Therefore, to lowest order we have to
solve the following two equations for shear and expansion

\be
 - \frac{d \sigma_{\mu \nu}}{d v} + \kappa \sigma_{\mu \nu} = \mathcal{E}_{\mu \nu}
\ee

and

\be
- \frac{d \theta}{dv} + \kappa \theta = S(v) + \sigma^2\, .
\ee

These equations are solved using advanced Green's function method and they are given by:
\be
 \label{sigmasol} \sigma_{\mu \nu}(v) = \int_{-\infty}^{\infty}
\mathcal{E}_{\mu \nu}(v') \ e^{\kappa (v-v')} \Theta(v'-v) dv'
\, ,\ee

\be
 \label{thetasol} \theta(v) = \int_{-\infty}^{\infty} \left( S(v') + \sigma^2(v') \right) e^{\kappa (v-v')} \Theta(v'-v) dv'.
\ee

So, our strategy will be to first consider the generators of the Rindler horizon which do not intersect the body, then the only source for expansion is
the shear generated by the conformal tensor via the tidal force equation. Next, we calculate the expansion from Eq.(\ref{thetasol}) and obtain the bound on
the size of the object using Eq.(\ref{condn}). The case of the generators which intersect the body will be treated separately. \\

\section{Charged object falling across RH}
\label{sec:RN}

Before we consider the rotating case, we first discuss the obvious and simplest generalization of \cite{AMV} and consider a charged spherical object
passing into the Rindler horizon. This case is almost similar to the uncharged case as the problem is still spherically symmetric.\\
To start with we adopt Minkowski coordinates $T, Z, x, y$ to describe Rindler space time. We assume the trajectory of the object to be

\be
 Z=z_0 \quad
\textrm{and}\,\, x=y=0
\,. \ee

The horizon of Rindler spacetime is the surface at $T=Z$. We will assume the mass $m$ and the charge of the object to be small so that it's influence can
be treated perturbatively.  As a result, we consider the object as a solution of linearized Einstein equation represented by a linearized
Reissner-Nordstrom metric. In the isotropic coordinates $(T,Z, \rho, \theta)$, the metric takes the following form:

\bea
ds^2 &=&
-\left(1-\frac{2 \, m}{\sqrt{\rho^2 + (Z-z_0)^2}}+\frac{q^2}{\rho^2 + (Z-z_0)^2} \right)\, dT^2\nonumber
\\&+& \left(1+\frac{2 \, m}{\sqrt{\rho^2 + (Z-z_0)^2}}-\frac{q^2}{2(\rho^2 + (Z-z_0)^2)} \right)\left(dZ^2+d\rho^2+\rho^2 d\theta^2 \right), \label{RN}
\eea

where $q$ is the charge of the object as measured by an asymptotic observer and  $\rho^2 = x^2+y^2$. We first treat the generators which don't cross the
matter. Due to spherical symmetry, we will only have the diagonal transverse components of the electric part of the Weyl tensor to be non-vanishing. We
will calculate these components on the horizon i.e. on the $T=Z$ surface. They are given by,

\be
\label{RN1}
  \mathcal{E}_{\rho \rho} = -\left( \frac{3 \rho^2 T^2 m  \kappa^2
}{(\rho^2\,+\,(Z-z_0)^2)^{5/2}}\,-\frac{3 q^2 \rho^2 T^2 \kappa^2}{(\rho^2 \, +\, (Z-z_0)^2)^3}\right) \, +\, O(mq^2)=
-\frac{1}{\rho^2} \mathcal{E}_{\theta \theta} \,.
\ee

On the horizon,

\be T=Z=z_0e^{\kappa \bar{v}}\,.\ee

Where the $\bar{v}=v-v_0$ is the shifted Killing time and $v_0$ is the time when the object hits the horizon. Now to get control over the time dependence of the
relations (\ref{RN1}), we will assume $\kappa\bv\ll1$ and approximate the components of the Weyl tensors as:
\bea
\label{RN2}
\mathcal{E}_{\rho \rho} &=& \mathcal{E}_{\rho \rho}^{(m)}\,+\mathcal{E}_{\rho \rho}^{(q)}\nonumber \\
   &=& -3 \,\rho^2\,z_{0}^2 \,\kappa^2\left( \frac{ m
}{(\rho^2\,+\,z_0^2\kappa^2\bv^2)^{5/2}}\,-\frac{ q^2}{(\rho^2 \, +\, z_0^2\kappa^2\bv^2)^3}\right) \, +\, O(mq^2)= -\frac{1}{\rho^2} \mathcal{E}_{\theta
\theta} \eea

\begin{figure}[h]
 \begin{center}
 \includegraphics[width=15cm,height=8cm]{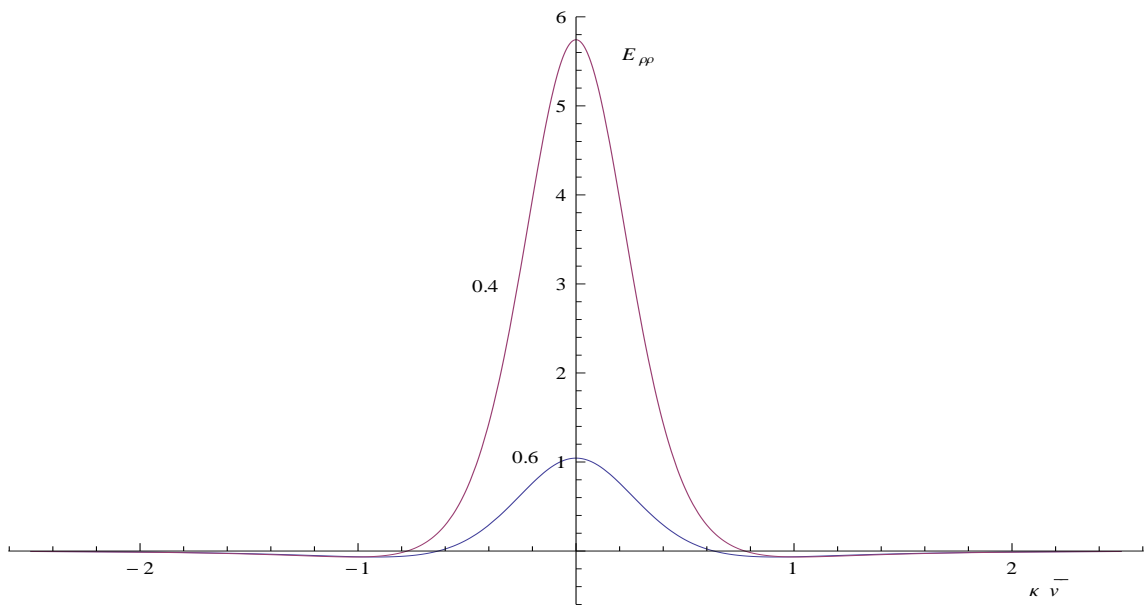}
 \end{center}
\caption{Plots of normalized $E_{\rho\rho}$ versus $\kappa \bv$ for two values of $\frac{\rho^2}{Z_0^2}$. It is clear from the plots that dominating contribution of the profile comes near $\kappa \bv=0$ and it gets better as we lower the ratio $\frac{\rho}{Z_0}$. Note that, one may approximate this as a delta function centered at $\bv =0$.} \label{RNplot}
\end{figure}

The components of the electric part of the Weyl tensor in (\ref{RN2}) have three extrema at $\bv=0$ and $\bv=\pm v_m=\pm \frac{1}{\kappa}\sqrt{\frac{36q^4}{25m^2z_0^2}-\frac{\rho^2}{z_0^2}}$. Our analysis will be valid in the region where the transverse distance of the object from the origin is very small compared to the height from which it is falling i.e. we always assume $\frac{\rho}{Z_0}\ll1$. It can be verified from (\ref{RN2}), $\mathcal{E}_{\rho\rho}$ has very negligible contribution around it's nonzero extremum points as we lower the ratio $\frac{\rho}{Z_0}$. This behavior of the conformal tensors is demonstrated in Fig.(\ref{RNplot}), where we have plotted the normalized field tensor $E_{\rho\rho}=\frac{z_0^4}{3q^2\rho^2\kappa^2}\mathcal{E}_{\rho \rho}^{(q)}-\frac{z_0^3}{3m\rho^2\kappa^2}\mathcal{E}_{\rho \rho}^{(m)}$ with $\kappa\bv$.

Since the extrema at $\bv=0$ is dominating we can ignore the contributions of the conformal tensors around other two extrema and approximate the tidal
profile as delta function peaked at $\bv=0$.

We now compute the tidal pulses explicitly. The approximated values of the non-vanishing components of conformal tensor are given by,

\be
 \mathcal{E}_{\rho \rho}(\bv) = \left(- \frac{4 m \kappa z_0 }{\rho^2}+\frac{9\pi\kappa q^2 z_0}{8 \rho^3}\right)
\delta(\bv) \, +\, O(mq^2)= -\frac{1}{\rho^2} \mathcal{E}_{\theta \theta} \, .\label{elecRN}
\ee

The corresponding expressions for shear components are now obtained utilizing (\ref{sigmasol}) as,

\be
\sigma_{\rho\rho}=-\frac{\sigma_{\theta\theta}}{\rho^2}=\left(- \frac{4 m \kappa z_0 }{\rho^2}+\frac{9\pi\kappa q^2 z_0}{8 \rho^3}\right)e^{\kappa \bv}
\Theta(-\bv)\,+\,O(mq^2)\, .
\label{RNsig}\ee

Now the expansion is obtained putting (\ref{RNsig}) into (\ref{thetasol}),

\be
\theta(\bv)=\int 2 \sigma_{\rho\rho}^2e^{\kappa(\bv-\bv')}\Theta(\bv'-\bv)d\bv'
\label{RNshear}.\ee

To first order in perturbation parameters, this yields the following expression for expansion,

\be \theta(\bv)=\frac{2}{\kappa}\left(\frac{-4m\kappa z_0 }{\rho^2}+\frac{9\pi\kappa z_0 q^2}{8\rho^3}\right)^2e^{\kappa \bv}(1-e^{\kappa \bv})\Theta(-\bv)
\label{expRN}\,.\ee

The condition for formation of caustics is now obtained with the help of (\ref{condn}),
\be \frac{\theta_{max}}{2\kappa}=\frac{1}{4\kappa^2}\left[\left(\frac{4m\kappa z_0 }{\rho^2}\,-\frac{9\pi q^2 \kappa z_0}{8\rho^3}\right)^2\right]\gtrsim 1
\label{RNcond}. \ee

Which for $q=0$ reduces to exactly the situation presented in \cite{AMV}. The bound on the object's size can now be obtained from
(\ref{RNcond}) setting $\rho = L$, the size of the object,

\be
\left(2mz_0\,-\, \frac{9\pi q^2 z_0}{16 L}\right)\geq L^2 \,\,\,\textrm{and}\,\,\, \left( \frac{9\pi q^2 z_0}{16 L}- 2mz_0 \right)\geq L^2
\label{modineq}\ee
Therefore, we get two inequalities satisfied by the object's size $L$ here. The first of these inequalities is given by,

\be  2mz_0- \frac{9\pi q^2 z_0}{16 L} \geq L^2 . \label{ineq1}\ee

To get the allowed values for $L$, we consider the corresponding cubic equation of the above inequality. This equation has the structure $ x^3 - a \, x + b = 0$ with $a>0$ and $b>0$. According to Descartes's rule of signs this equation will always have at least one positive root and that root essentially gives the lower bound on the size of the charged object so that no caustic forms to the future of the bifurcation surface. The solution of the equation corresponding to (\ref{ineq1}) may be expressed as small departure from the solution for the uncharged case,

\be L^2 \approx 2\, m \, z_0\,-\,\frac{9\pi z_0 q^2}{16\sqrt{2 mz_0}} \,+\,O\left(\frac{q^4}{m^2}\right) \label{RNcond1}\, .\ee

Hence the area of the object should be larger than the rhs of (\ref{RNcond1}) to validate the PPFL in the desired region and we get the following condition on the object's surface area,

\be L^2 \geq 2\, m \, z_0\,-\,\frac{9\pi z_0 q^2}{16\sqrt{2 mz_0}} \,+\,O\left(\frac{q^4}{m^2}\right) \label{RNcond2}\, .\ee

Also, the condition (\ref{RNcond2}) shows that the lower bound on the size of the object tends to the desired value when charge $q$ is taken to be zero \cite{AMV}. The second inequality which we get from the condition (\ref{modineq}) is,

\be
L^2 \leq \frac{9\pi q^2 z_0}{16 L}-2mz_0 
\label{ineq2}\, .\ee

Note that, this is obtained only when, $m < \left( 9 \, \pi q^2/ 32 L\right)$. Although the equation corresponding to (\ref{ineq2}) will also have a real positive solution but that solution can not be related to the one for neutral case as we are not allowed to take $q \rightarrow 0$ limit anymore. Hence, we don't consider this branch of solution. Thus for all practical purposes, the condition (\ref{RNcond2}) ensures no formation of caustic in the future of the bifurcation surface and the second order terms in the Raychaudhuri equation can be ignored.

Now, for generators crossing the object we may follow the arguments presented in  \cite{AMV} as both the cases are spherically symmetric. For charged object, if we assume that the object is of homogeneous energy density, then again the electric part of the Weyl tensor inside the object will be smaller than outside and this implies that the condition (\ref{RNcond2}) will still remain sufficient to rule out any caustics in the region of our interest.

\section{Slowly rotating object falling across RH}
\label{sec:Kerr}

We now study the tidal distortion caused by a weakly rotating self-gravitating object freely falling across the Rindler horizon. We need to solve the
equations (\ref{sigmasol}) and (\ref{thetasol}). We adopt the same coordinate system as described in the previous section. The object is falling from the
same position at $Z=z_0$ and following the same trajectory like the charged case. We will again assume the mass $m$ of the particle to be small so that
it's influence can be treated perturbatively.

 Now, for small rotation we can approximate the object to be linearized Kerr metric. In the cylindrical coordinates for the
 transverse spatial part of the object we can cast the linearized
 Kerr metric in the quasi-isotropic form \cite{cohen, Lanza},

\bea ds^2=&-\left(1-\frac{2m}{\sqrt{\rho^2 + (Z-z_0)^2}}\right)dT^2-\frac{4m\,a \rho^2}{(\sqrt{\rho^2 + (Z-z_0)^2})^3}\,d\theta dT \nonumber \\& +
\left(1+\frac{2m}{\sqrt{\rho^2 + (Z-z_0)^2}}\right)(d\rho^2+dZ^2+\rho^2d\theta^2)\label{kerrmtric} \, ,\eea

where $a$ is the angular momentum per unit mass of the object. The non vanishing components of the Electric part of Weyl tensor corresponding to this
metric are

\be \mathcal{E}_{\rho\rho}=-\frac{\mathcal{E}_{\theta\theta}}{g_{\theta\theta}}=-\frac{3 m\kappa^2\rho^2 T^2}{(\rho^2 + (Z-z_0)^2)^{5/2}}\,+ O(m^2)
\label{Erho}\ee and

\be \mathcal{E}_{\rho\theta}=-\frac{15 TZ (Z-z_0) ma \kappa^2 \rho^3}{(\rho^2 + (Z-z_0)^2)^{7/2}}\, + O(m^2a) \label{Ertheta}\, .\ee

As in the previous case, we first consider the generators which do not cross the object. We again wish to employ the delta function approximation to manage
the time dependence of the relations (\ref{Erho}) and (\ref{Ertheta}) and assume $\kappa\bv$ to be small. In the lowest order, the diagonal transverse
parts of electric part of Weyl tensor ( i.e. $ \mathcal{E}_{\rho\rho}$ and $\mathcal{E}_{\theta \theta}$) remains the same and can be approximated as a
delta function exactly in same way as in the non rotating case given by,

\be
 \label{delta1} \mathcal{E}_{\rho \rho}(\bv) = - \frac{4 m \kappa z_0 }{\rho^2} \delta(\bv) = -\frac{1}{\rho^2} \mathcal{E}_{\theta \theta} \,.
 \ee

The mixed component $\mathcal{E}_{\rho\theta}$ has to be treated carefully. If we assume $\kappa\bv$ to be small, this component can be approximately
written as

\be \mathcal{E}_{\rho\theta} = -\frac{15 z_0^{3} \kappa^3 \rho^3 \bv m a}{(\rho^2 + z_0^2 \kappa^2 \bv^2)^{7/2}} \,.\ee

Note that, unlike the diagonal component, $\mathcal{E}_{\rho\theta}$ vanishes at $\bv = 0$ but also has two extremum points $\bv = \pm v_m$. As a result,
it can no longer be approximated as only a delta function source centered at $\bv = 0$. We want to emphasize a point here, although the object hits the horizon at $v=v_0$ but it turns out that the whole process can now be modeled as three tidal pulses hitting the horizon at three different times. While the diagonal component of Weyl tensor hits the horizon at $v=v_0$, the off-diagonal parts hit it at different times $v=v_0\pm v_m$ (as shown bellow). So, we now represent the off-diagonal components of conformal tensors by two separate delta functions around
two extrema and write those as

 \be
 \label{delta2} \mathcal{E}_{\rho\theta}^{\pm}(\bv)=(\mp) \frac{3 ma \kappa z_0 }{\rho^2} \delta(\bv \pm
v_m)\, ,
 \ee

 where
\[v_m=(\pm)\frac{\rho}{\sqrt{6}\kappa z_0}\]
are the points at which $\mathcal{E}_{\rho\theta}$ is sharply peaked. Note that this approximation is valid in the region where $\rho \ll z_0$ as it is
consistent with our assumption that the particle is weakly gravitating. Also, the mixed component $\mathcal{E}_{\rho\theta}$ changes it's sign around $\bv
=0$ and therefore the shear due to the conformal tensor also changes it's sign at $\bv = 0$. But, since the expansion is sensitive to the square of the
shear, the horizon indeed expand all the time in accordance to Hawking's area increase theorem.\\

We now compute the shear components with the help of (\ref{sigmasol}) to get,

\begin{figure}[h]
 \begin{center}
 \includegraphics[width=10cm,height=6cm]{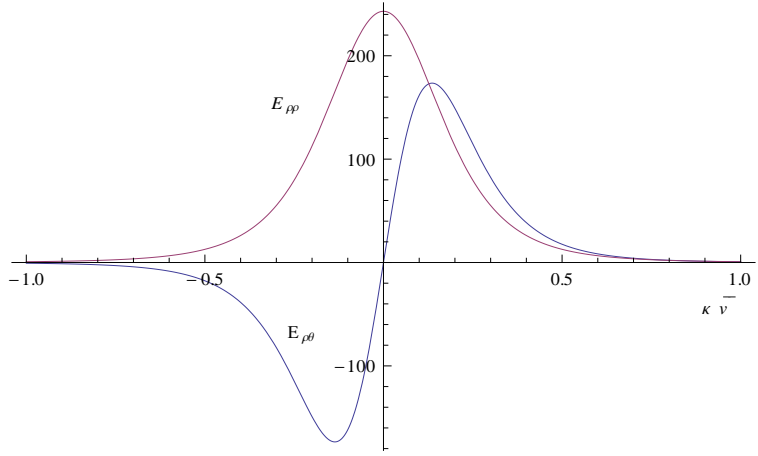}
 \end{center}
\caption{The profile of $E_{\rho\rho}=-\frac{z_0^3}{3m\rho^2\kappa^2}\mathcal{E}_{\rho\rho}$ and $E_{\rho\theta}=-\frac{z_0^4}{15 \rho^3 \kappa^2 ma}\mathcal{E}_{\rho\theta}$ for the slowly rotating case with $\frac{\rho}{z_0}={1 \over 4}$}\label{plot2}
\end{figure}

\be \sigma_{\rho \rho} =-\frac{\sigma_{\theta\theta}}{\rho^2}=- \frac{4 m \kappa z_0 }{\rho^2}e^{\kappa \bv}\Theta(-\bv)  \label{sig1} \ee and \be
\sigma_{\rho \theta}^{\pm}=(\mp)\frac{3 ma \kappa z_0 }{\rho^2}e^{\kappa (\bv \mp v_m)}\Theta(\pm v_m- \bv) \label{sig2} .\ee We now put (\ref{sig1}) and
(\ref{sig2}) into (\ref{thetasol}) and compute the expansion for these tidal pulses.

\be \theta(\bv)=I_0\,+\,I_{+}\,+\,I_{-}\label{expan}\,.\ee Where \be
I_0=\int_{-\infty}^{\infty}\left[2\sigma_{\rho\rho}^2\,e^{\kappa(\bv-\bv')}\Theta(\bv^{'}-\bv)\right]d\bv^{'} , \label{0}\ee 
\be
I_{+}=\int_{-\infty}^{\infty} \left(\frac{\sigma_{\rho \theta}^{+}}{\rho}\right)^2\,e^{\kappa(\bv-\bv^{'})}\Theta(\bv^{'}-\bv)d\bv^{'} \label{1} \ee and
\be I_{-}=\int_{-\infty}^{\infty} \left(\frac{\sigma_{\rho \theta}^{-}}{\rho}\right)^2e^{\kappa(\bv-\bv^{'})}\Theta(\bv^{'}-\bv)d\bv^{'} \label{2}.\ee 
The
expansion (\ref{expan}) is now sum of three sources peaked at three different Killing times. The $I_0$ integral only contains the mass of the in-falling
matter and the other two integrals $I_{\pm}$ contains the angular momentum dependent part. Integrating (\ref{0}), (\ref{1}) and (\ref{2}) we get, 

\bea
\theta(\bv)&=&\frac{2}{\kappa}\left(\frac{-4m\kappa z_0 }{\rho^2}\right)^2\,e^{\kappa \bv}(1-e^{\kappa \bv})\Theta(-\bv) \nonumber \\&+&
\frac{1}{\kappa}\left(\frac{-3 ma \kappa z_0 }{\rho^3}\right)^2\,e^{\kappa (\bv - v_m)}(1-e^{\kappa (\bv-v_m)})\Theta(v_m-\bv)\nonumber
\\&+&\frac{1}{\kappa}\left(\frac{3 ma \kappa z_0 }{\rho^3}\right)^2\,e^{\kappa (\bv + v_m)}(1-e^{\kappa (\bv + v_m)})\Theta(-v_m-\bv). \eea 

The maximum value of the expansion is given by,

\bea
\theta_{max}&=&\frac{2}{\kappa}\left(\frac{-4m\kappa z_0 }{\rho^2}\right)^2\,\times  \left({1\over 4}\right)\,+\,
\frac{1}{\kappa}\left(\frac{-3 ma \kappa z_0 }{\rho^3}\right)^2\,\frac{e^{-\kappa v_m}}{2}\left(1-{1\over 2}e^{-\kappa v_m}\right)\nonumber
\\&+&\frac{1}{\kappa}\left(\frac{3 ma \kappa z_0 }{\rho^3}\right)^2\,\frac{e^{\kappa v_m}}{2}\left(1-{1\over 2}e^{\kappa v_m}\right).\eea

Since $\kappa v_m (\approx \frac{\rho}{z_0}\ll 1)$ is small we can expand $\theta_{max}$ in a Taylor series to get, 

\be
\theta_{max}=\frac{1}{2\kappa}\left[\left(\frac{4m\kappa z_0 }{\rho^2}\right)^2\,+\,\left(\frac{3 ma \kappa z_0}{\rho^3}\right)^2
\,+\, O((\kappa v_m)^2)\right]
.\ee

Therefore, sub leading terms in $\theta_{max}$ is proportional to $\left(\frac{\rho}{z_0}\right)^2$ can be ignored for the rest of this section. Now the remaining task is to calculate the bound on the size of the object. Using  (\ref{condn}), we get the condition for formation of caustic to the future of the bifurcation surface, represented by the following inequality,

\be
\frac{1}{4\kappa^2}\left[\left(\frac{4m\kappa z_0 }{\rho^2}\right)^2\,+\left(\frac{3 ma \kappa z_0}{\rho^3}\right)^2\right]\gtrsim 1
\label{causticcnd}.\ee

The transverse extension of the object is captured by the coordinate $\rho$ which we now set to be equal to $L$, the size of the object. The condition (\ref{causticcnd}) can now be recast as following inequality in $L$,

\be
4L^6-16m^2z_0^2L^2-9m^2a^2z_0^2 \lesssim 0
\label{hexaineq}\,.\ee

The solution of the corresponding equation of the inequality (\ref{hexaineq}) will give the upper bound on the size of the object below which there will always be caustics and PPFL will be violated. Employing Descartes's rule of sign we can easily verify that indeed there exists exactly one positive real root for this equation. Therefore caustics will never form if the object size $L$ is greater than this positive root. However here we don't attempt to solve the sixth order equation as it is rather difficult to extract the positive root instead we obtain a stronger condition on $L$ by solving another equivalent inequality from (\ref{causticcnd}). For positive values of $L$ we can rewrite the inequality (\ref{causticcnd}) as,

\be
\left[\left(\frac{2m z_0 }{L^2}\,+\frac{3 ma z_0}{2L^3}\right)^2\right] \gtrsim 1
\label{nwcnd}\ee

where we have ignored a positive definite term which is twice the product of the two terms in the parenthesis in the rhs of (\ref{nwcnd}). Now this inequality will lead to the following cubic inequality in $L$,

\be
L^3-2mz_0L-{3 \over 2}m a z_0 \lesssim 0
\label{cubic3}\,.\ee

The positive root of the equation corresponding to (\ref{cubic3}) can be expressed as small departure from the root without the rotation parameter $a$ as earlier case,

\be
L^2 \approx 2mz_0 \,+ \, \frac{3\sqrt{mz_0}a }{2\sqrt{2}}\, +\,O(a^2)
\label{bound}.\ee

In the region above this root one can easily verify the polynomial is increasing
and therefore the allowed value for area of the rotating object such that formation of caustics in the region of interest can be avoided is given by 

\be
L^2 \geq 2mz_0 \,+ \, \frac{3\sqrt{mz_0}a }{2\sqrt{2}}\, +\,O(a^2)
\label{allowed}.\ee

As we expect, the expression (\ref{allowed}) reduces to exactly the corresponding condition for non-rotating case described in \cite{AMV} when the angular momentum parameter $a$ set to zero. Note that the region allowed by (\ref{allowed}) is obtained ignoring a positive definite term from the original inequality (\ref{causticcnd}) which means there may be still a value of $L$ lower than this beyond which we can still have the condition (\ref{causticcnd}) to hold. So this is actually a stronger condition for the validity of PPFL. But as long as the characteristic size of the object obeys the following condition, it is guaranteed there will be no caustic formation in the future of the bifurcation surface and the derivation of PPFL will hold.

This ends the analysis for generators which are not crossing the matter. For generators which crosses the in-falling object the situation is rather uncertain as no interior solution for Kerr space time is known till date. Therefore, we cannot analyze the situation similarly as in the non-rotating cases. We can only say, if the conformal tensor inside the rotating object is smaller than outside then the condition (\ref{allowed}) will still be sufficient to discard any formation of caustics to the future of the bifurcation surface, although there is no scope to verify this right now.  

\section{Discussions}
The PPFL has been examined in the context of Rindler horizon for general processes when a rotating or a charged object crosses the horizon. It has been established that for sufficiently quasi-stationary process we can avoid formation of caustics to the future of bifurcation surface of the background stationary space time. The condition of non-formation of caustics can be estimated through a threshold value of size of the object. Although not mentioned explicitly, this analysis is valid for any horizon which can be approximated as a bifurcate Killing horizon so long the curvature scale $l$ of the ambient space-time satisfies $l\gg r$ \cite{AMV}. This is because one has the freedom of choosing the past and future limit of the integrations ($v=$constant surfaces) to be such that this interval is much larger compared to the characteristic response time ($\kappa^{-1}$) of any dynamical process happening across  the horizon.

In \cite{AMV}, the PPFL holds when the change in area or entropy induced due to the passage of the non rotating and uncharged object (of Killing energy $E_{\chi}$) is much smaller than the change in area of the horizon ($\sim L^2$). This is quantitatively expressed as, 

\be
L^2 > E_{\chi}/\kappa
\ee

However, in case of the charged object, we may express the condition (\ref{RNcond2}) as

\be
L^2 \sim 2 \varepsilon/\kappa,
\label{intp}\ee
where 
\be
\varepsilon=\kappa z_0\left(m-\frac{9\pi q^2}{32 \sqrt{2z_0m}}\right)
\ee
is some effective energy associated with the object induced by the process. Similarly for the rotating case from (\ref{allowed}), we can write a condition like (\ref{intp}) with 
\be
\varepsilon=\kappa z_0\left(m + \, \frac{3\sqrt{m}a }{4\sqrt{2z_0}}\right)
\ee
If we use the interpretation in \cite{AMV}, we may like to interpret the expressions of $\varepsilon$ as the Killing energy imparted on the horizon due to the passage of the object.\\
The analysis presented here for charged object can easily be generalized to any $d\geq 4$ dimension. However generalizing for rotating case needs more careful analysis as the solutions may not uniquely belong to Kerr family.\\
In conclusion, our work generalizes the result in \cite{AMV} incorporating the effect of both charge and rotation of the perturbing object. In accordance with \cite{AMV}, we also infer that the derivation physical process law for Rindler horizon is well defined and it is possible to use quasi stationary approximation consistently provided the size of the perturbing object obeys certain condition.

\section*{Acknowledgements}
We thank K. P. Yogendran and Sanved Kolekar for useful discussion. The research of SS is partially supported by IIT Gandhinagar start up grant no. IP/IITGN/PHY/SS/201415-12. SS also acknowledge the warm hospitality of the physics theory group at University de Santiago de Compostela, Spain where part of this work is done.

\end{document}